\documentclass[aps,pra,showpacs,amsmath,amssymb,lengthcheck]{revtex4}

\usepackage{epsfig}
\begin{document}
\title{Collective Excitations of Trapped Imbalanced Fermion Gases}
\author{Achilleas Lazarides and Bert Van Schaeybroeck}
\affiliation{Instituut voor Theoretische Fysica, Katholieke Universiteit Leuven,\\ Celestijnenlaan
  200 D, B-3001 Leuven, Belgium.}

\begin{abstract}
We present a theoretical study of the collective excitations of a
trapped imbalanced fermion gas at unitarity, when the system
consists of a superfluid core and a normal outer shell. We
formulate the relevant boundary conditions and treat the normal
shell both hydrodynamically and collisionlessly. For an isotropic
trap, we calculate the mode frequencies as a function of trap
polarization. Out-of-phase modes with frequencies below the
trapping frequency are obtained for the case of a hydrodynamic
normal shell. For the collisionless case, we calculate the
monopole mode frequencies, and find that all but the lowest mode
may be damped.
\end{abstract}

\maketitle The recent pioneering experiments on polarized fermion
systems in the BEC-BCS crossover make a broad range of exciting
new phenomena accessible and offer the opportunity to study
longstanding issues in the field of strongly interacting many-body
systems~\cite{zwierlein,partridge}. By loading a trap with unequal
numbers of particles of two spin states, a phase-separated system
consisting of an unpolarized superfluid (SF) core and a
surrounding polarized normal (N) phase is obtained. Experimentally
probing the collective excitation frequencies in ultracold gases
is, by now, a standard technique, and has proved indispensable to
understanding such systems. An experimental study of collective
modes in imbalanced fermionic superfluid systems in the near
future seems, therefore, a realistic prospect.

The N-SF interface plays a major role in the static properties of the
polarized system: taking into account the interface tension has proved
essential to explaining the experiments performed in highly anisotropic
traps~\cite{desilva,haque}. Moreover, its presence affects the thermal
equilibration process at low temperatures~\cite{vanschaeybroeck} and has been
speculated to cause large fluctuations in the polarization at the
interface~\cite{zhai}.

In this Rapid Communication, we study the effects of polarization on the
frequencies of the collective excitations. To do this, it is necessary first
to establish the boundary conditions encoding the important physics at the
interface. We then study two distinct cases: in one, the normal part is
assumed to behave hydrodynamically, and in the other, collisionlessly, so that
it is described by a Boltzmann-Vlasov equation. We present results for
isotropic traps only; results for a highly elongated trap will be presented
separately, along with more detailed calculations~\cite{lazarides}.

\begin{figure}[b]
   \epsfig{figure=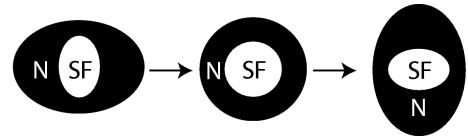,angle=0, height=50pt}
   \caption{Time evolution of the quadrupole  ($\ell=2$) out-of-phase mode of
     an imbalanced fermion system in the hydrodynamic-normal-gas case".
     The outer boundaries of the normal (N) and the superfluid (SF) phase
     perform the same motion but $\pi$ out of phase.
   \label{achilleasshouldnotgoonholidays}
 }
 \end{figure}
For a fully hydrodynamic trapped system, we find two general
classes of excitations: A first class consists of modes which for
zero polarization reduce to the well-known single-component-system
excitations; for finite polarization, these evolve to modes with
the same general character of motion but slightly shifted
frequencies. In what follows, we refer to them as in-phase (IP)
modes, for reasons that will become clear. Additionally,
excitations unique to the two-component system  constitute a second
class. Particularly interesting is that, at low polarization,
these new modes have frequencies below the trapping frequency.
Each of these low-frequency excitations is in one-to-one
correspondence with the nodeless ($n=0$, $\ell$) modes of the
single-component system, to which their motion is related as
follows: the outer boundary of the N and SF components oscillate
in the same way as the corresponding one-component $n=0$ mode,
except that they are out of phase with each other. This is
illustrated in Fig.~\ref{achilleasshouldnotgoonholidays} for the
quadrupole mode.  For brevity, we refer to these new type of
excitations as the out-of-phase (OOP) modes. 

In the collisionless case, we study the monopole ($\ell=0$) modes.
Simultaneously solving the hydrodynamic and Boltzmann-Vlasov equations for the
SF and N phase respectively, and imposing the appropriate boundary conditions,
we find that the lowest, $n=1$, monopole mode remains at $2\omega_0$ for all
polarizations (here $\omega_0$ is the trapping frequency), in agreement with
an exact result obtained by Castin~\cite{castin}. On varying the polarization,
the $n=2$ monopole mode frequency crosses over from the value appropriate for
a fully-superfluid at zero polarization and a fully-collisionless system at
complete polarization.  For this (and higher) monopole modes, and for some
intermediate polarizations, damping occurs, the cause of which is purely
geometric.

We shall first describe the equilibrium state of the system, and then
separately present our calculations and results for the hydrodynamic and
collisionless cases.

\textit{Equilibrium} --- Consider a spin mixture of polarization
$(N_\uparrow-N_\downarrow)/(N_\uparrow+N_\downarrow)$ with
$N_\uparrow$ and $N_\downarrow$ the spin up and down particle
numbers, trapped by a harmonic potential $V(\mathbf{r})$. A
nonzero imbalance generally results in the appearance of at least
two phases: the unpolarized SF and the partially or completely polarized N
phase which consists, in general, of a mixture of $\downarrow$ and
$\uparrow$ particles.

When the N and SF phases are separated by a first-order phase
transition, an interface is formed between them with the SF in the
trap center.  Due to the universal nature of the strongly
interacting regime, the equilibrium pressure $\overline P$ and
density $\overline \rho$ in the SF are known~\cite{carlson} to be
related by $\overline
P_{SF}=\hslash^2\xi(3\pi^2)^{2/3}\overline\rho_{SF}^{5/3}/5m$,
with the universal constant $\xi\approx 0.41$ and $m$ the particle
mass; in the N phase, on the other hand, one has $\overline
P_{i}=\hslash^2(6\pi^2)^{2/3}\overline\rho_{i}^{5/3}/5m$ for
$i=\uparrow,\,\downarrow$ (the overline denotes
equilibrium values). Moreover, within the local density
approximation all three of $\uparrow$, $\downarrow$ and SF may be
described by an effective equilibrium chemical potential
$\overline \mu_j(\mathbf{r})=\mu^0_j-V(\mathbf{r})$, where
$\mu^0_j$ is the chemical potential at the center of the trap and
$j=\uparrow,\downarrow$ or SF~\footnote{The interface tension
  accounts for the most important corrections to the local density
  approximation~\cite{haque}.}.

Next we include the interface into our equilibrium framework. At zero
temperature, the SF is theoretically predicted to be unpolarized;
experimentally, the trap core is observed to be unpolarized. This
implies a negligible population of single-particle excitations in
the trap center at experimental temperatures.
Denoting the equilibrium position of the interface by $\overline\zeta$, chemical
equilibrium is then ensured by imposing~\cite{giorgini}
\begin{equation}\label{achilleasisamonkey}
  \left.2\overline\mu_{SF}\right\vert_{\overline\zeta}=\left(\overline\mu_{\uparrow}
    +\overline\mu_{\downarrow}\right)_{\overline\zeta}.
\end{equation}
Furthermore, for mechanical stability, the Laplace condition must be satisfied:
\begin{align}\label{achilleasisadonkey}
  \left(\overline P_{SF}-\overline P_{\uparrow}-\overline
    P_{\downarrow}\right)_{\overline\zeta}
  =\left. \overline{\sigma}\left(\frac{1}{R_1}+\frac{1}{R_2}\right)\right|_{\overline\zeta}.
\end{align}
Here, $R_1$ and $R_2$ are the radii of curvature of the interface, the
position of which is denoted by $\zeta$, and $\sigma$ are the interface tension~\cite{desilva,haque};
following Haque and Stoof~\cite{haque}, we take
$\sigma=0.6m\mu_{SF}^2/\hslash^2$~\footnote{Unitarity allows us to
write
  $\sigma=\eta m \mu_{SF}^2/\hslash^2$, with $\eta$ a dimensionless quantity
  depending on the local interface curvature. Haque and Stoof find $\eta=0.6$,
  independently of the polarization, therefore independently also of
  the curvature~\cite{haque}.}. Note that $\mu_{SF}$ is the total chemical
potential, including its fluctuation during an oscillation, which
reduces to $\overline\mu_{SF}$ at equilibrium.
Equations~\eqref{achilleasisamonkey}
and~\eqref{achilleasisadonkey}, together with the expressions for
the densities and pressures, allow us to fix the position of the
interface for given particle number and polarization in the global
equilibrium state. To do this, all that is necessary is to solve
Eqs.~\eqref{achilleasisamonkey} and~\eqref{achilleasisadonkey} for
the radii while imposing the total particle number to be $7\times
10^6$. We then find that, for all polarizations, the superfluid
core is surrounded by a fully polarized normal shell of the
majority species. This is similar to what is observed in the Rice
experiments~\cite{partridge,haque}.

Dynamical phenomena, such as the collective excitations studied here, involve
situations in which the local velocities are nonzero. For such cases,
involving departures from the global equilibrium state, we must extend
Eqs.~\eqref{achilleasisamonkey} and~\eqref{achilleasisadonkey} appropriately
as well as supplement them with additional boundary conditions.

\textit{Dynamics: Hydrodynamic Normal Gas} --- Consider first the
case in which the normal phase behaves hydrodynamically, that is,
when inter-particle collisions are frequent enough for local
thermodynamic equilibrium to be ensured everywhere during the
oscillation. The chemical potential and pressure remain
well-defined quantities and therefore
Eqs.~\eqref{achilleasisamonkey} and~\eqref{achilleasisadonkey},
with $\overline\mu$ and $\overline P$ replaced by
$\overline{\mu}+\delta \mu$ and $\overline{P}+\delta P$,
respectively, remain applicable.

In the system under study, particles may pass over from the N to
the SF phase and vice versa~\cite{vanschaeybroeck}. This
interconversion separately conserves the mass of each species.
Denoting the position of the interface by
$\zeta=\overline\zeta+\delta \zeta$, where $\delta\zeta$ is the
departure of the interface from the equilibrium position, the
appropriate boundary conditions (one for each of
$i=\uparrow,\,\downarrow$) are, at each point on the interface,
\begin{align}\label{achilleasshouldbehave}
  \mathbf{e}_\zeta\cdot\left(2\mathbf{v}_{i}\rho_{i}
    -\mathbf{v}_{SF}\rho_{SF}\right)_{\zeta}
  =(2\rho_{i}-\rho_{SF})_\zeta\partial_t\delta\zeta,
\end{align}
in which $\boldsymbol{e}_\zeta$ is the unit vector perpendicular
to the interface (and directed towards the N side) and
$\mathbf{v}$ are the velocities. These boundary conditions are,
essentially, a restatement of the continuity equation for the case
of two coexisting phases of different densities with a movable,
\textit{permeable} interface separating them.

\begin{figure}
  \epsfig{figure=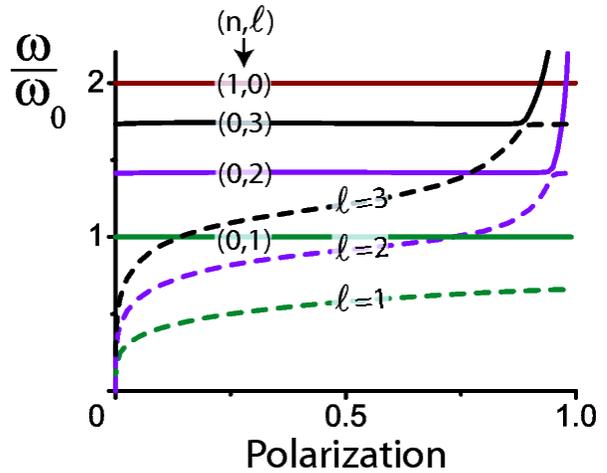,angle=0, height=180pt}
  \caption{Collective mode frequencies against polarization for a trap with a hydrodynamic normal shell and with $7\times
    10^6$ particles. The dashed and full lines denote the out-of-phase and in-phase collective modes
    respectively. The frequencies of the in-phase modes are rather insensitive
    to the polarization, except for polarizations after the avoided
    crossing. At zero polarization, the IP modes reduce to
     the excitations of the single-component systems, labelled by the numbers $n$ and $\ell$
    \label{fig_achilleasshouldcleanthehouse}
  }
\end{figure}
Finally, the bulk dynamics for both the SF and the N are governed by the Euler
and continuity equations. Applying standard techniques~\cite{pethick}, we
linearize these and obtain the equation for the deviation of the chemical
potential from its equilibrium value, $\delta\mu$. Taking the temporal and
angular parts of $\delta\mu_{SF}$, $\delta\mu_\uparrow$,
$\delta\mu_\downarrow$, and $\delta\zeta$ to be $e^{i\omega t}$ and $Y_l^m$
respectively, with $Y_l^m$ the spherical harmonics, we obtain
$\delta\mu_{SF}\propto
r^\ell\text{F}\left(\alpha^+,\alpha^-,\alpha^0,(r/R_{SF})^2\right)$ and
$\delta\mu_{i}\propto
r^\ell\text{F}\left(\alpha^+,\alpha^-,3/2,1-(r/R_{i})^2\right)$ with
$i=\uparrow,\downarrow$ and $\text{F}$ the hypergeometric function. Here
$\alpha^0=\ell+3/2$, $2\alpha^\pm=\ell+2\pm[\ell^2
+\ell+4+3\omega^2/\omega_0^2]^{1/2}$ and $\ell$ a positive integer or zero.
For all three phases, we have also defined $R^2_j\equiv 2\mu^0_j/m\omega_0^2$
where $j=\uparrow,\downarrow$ or SF. The linearized versions of
Eqns.~\eqref{achilleasisamonkey},~\eqref{achilleasisadonkey}
and~\eqref{achilleasshouldbehave}, together with the linearized bulk
solutions, now suffice to describe the dynamical behavior of the system in the
case where the normal side is hydrodynamic~\cite{lazarides}.

\textit{Results: Hydrodynamic Normal Gas} ---  In
Fig.~\ref{fig_achilleasshouldcleanthehouse} we show the dependence
of the mode frequencies on the polarization. We distinguish
between the two kinds of excitations already mentioned: the
in-phase (IP) modes, analogous to the excitations present in
single-component systems (full lines), and the out-of-phase (OOP)
modes, unique to two-component systems (dashed lines). Note that
similar excitations appear in trapped boson-boson
systems~\cite{svidzinsky}. At zero polarization, the IP mode
frequencies reduce to the single-component frequencies
$\omega^2/\omega_0^2=\ell+2n[(2/3)(n+\ell+1/2)+1]$~\cite{heiselberg}
which are in almost exact agreement with the
experiments~\cite{giorgini}. For all polarizations, the frequency
of the sloshing (or Kohn) mode remains exactly $\omega=\omega_0$;
this excitation corresponds to a rigid cloud motion and is
therefore not affected by the bulk equation of state. The
frequencies of the IP modes with $n=0$ and $\ell>1$, on the other
hand, vary slightly with polarization; this is a consequence of
the nonzero interface tension. We note here that recent
experiments at MIT reveal the existence of a large partially
polarized shell resulting from the interactions in the N phase.
The incorporation of such interactions is beyond the scope of this
work. Nevertheless, our main result, the existence of the OOP
modes, remains valid there since such modes are generic to the
two-component system~\cite{svidzinsky}.

\textit{Dynamics: Collisionless Normal Gas} --- At temperatures
well below the Fermi energy, a normal Fermi gas is expected to
behave collisionlessly, especially in the case when this phase is
fully spin polarized. We therefore turn our attention to a
collisionless normal gas, that is, one that is described by a
Boltzmann-Vlasov equation, as used, for example, to study a
(spatially mixed) Bose-Fermi system in Ref.~\onlinecite{maruyama}.
The issue of collective excitations in normal fermion gases has
been addressed using several methods which include the
hydrodynamic approximation, the method of averaging, the scaling
ansatz method, the sum-rule approach, and the random-phase
approximation~\cite{bruun2}.

The fully polarized normal gas is now described by a distribution function
$f(\mathbf{r},\mathbf{v},t)$, evolving according to the Boltzmann-Vlasov
equation
\begin{equation}\label{eq:vlasov}
\partial_tf+\boldsymbol{v}\cdot\boldsymbol{\nabla}_r
f-\omega_0^2\mathbf{r}\cdot\boldsymbol{\nabla}_v f=0,
\end{equation}
rather than the hydrodynamic equations. We shall again consider small
deviations from the equilibrium function, taking $f$ to deviate from the
equilibrium value by a non-isotropic deformation of the Fermi
surface~\cite{lifshitz}:
$f(\mathbf{r},\mathbf{p},t)=f_0(\mathbf{r},\mathbf{p},t)+\delta(|\mathbf{v}|-v_F)\nu(\mathbf{r},\mathbf{p},t)$,
with $f_0$ the Fermi function and $v_F$ the (position-dependent) Fermi
velocity. The interesting question then arises of which boundary conditions
apply at an interface between a hydrodynamic and collisionless
gas~\cite{giorgini}. First, notice that, since we no longer have local
thermodynamic equilibrium, there is no analogue of
Eq.~\eqref{achilleasisamonkey} for the deviations from the equilibrium
configuration. On the other hand, the Laplace condition,
Eq.~\eqref{achilleasisadonkey} remains valid with the following modifications:
a) since the normal gas is fully polarized, only quantities pertaining to the
majority species appear in it; b) the role of the pressure fluctuation of the
majority species, $\delta P_\uparrow$ in the previous section, is now played
by the radial component of the momentum flux tensor
$\delta\Pi_{rr}^\uparrow=m^4\int\text{d}^3\boldsymbol{v}\, v_r^2
(f-f_0)/(2\pi\hbar)^{3}$.

Next, from Eq.~\eqref{achilleasshouldbehave}, we see that the
radial component of the SF velocity must be equal to the radial
velocity of the interface, that is,
$\mathbf{e}_\zeta\cdot\mathbf{v}_{SF}=\partial_t\delta\zeta$, so
that there is no flux through the interface. There are therefore
only two possibilities for a particle incoming on the interface
from the N side: Andreev reflection or specular
reflection~\cite{vanschaeybroeck}. Here, Andreev reflection is
suppressed by the full spin polarisation in the N phase, leaving
specular reflection as the sole reflection mechanism. According to
Bekharevich and Khalatnikov, specular reflection off a moving
interface results in the boundary condition
$\nu(-\chi)=\nu(\chi)-2i \omega\chi\delta\zeta$ where
$\chi=\cos\phi$ and $\phi$ the angle between $\mathbf{r}$ and
$\mathbf{v}$.
 \begin{figure}
\epsfig{figure=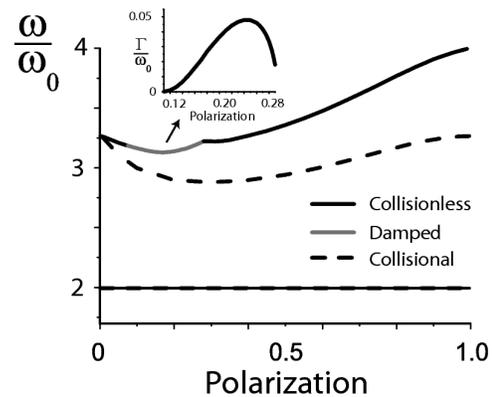,angle=0, width=180pt} \caption{Monopole
frequencies against polarization for a trap with $7\times 10^6$
particles. The lowest mode frequencies are exactly at $2\omega_0$,
both for the cases of a collisionless (full line) and hydrodynamic
(dashed line) normal shell. The second monopole frequencies
deviate substantially between these case. The grey full line
indicates the region of damping for the collisionless case and the
damping rates are shown in the inset.\label{spheric}
  }
\end{figure}
At $r=\overline\zeta$ the rotationally symmetric ($\ell=0$)
solution satisfying Eqn.~\eqref{eq:vlasov} and this boundary
equation is
\begin{align}\label{eq:soln}
\nu(\overline\zeta,\chi,t)=\chi\omega\delta\zeta[\cot(\omega\tau/2)+i],
\end{align}
where $\tau$ is the time for a classical particle of velocity
$\mathbf{v}$ to travel from the interface and back~\footnote{We
find
$\tau=\arctan[2\chi/(\widetilde{r}/\zeta-\zeta/\widetilde{r})]/\omega_0$
and $\widetilde{r}^2=R_\uparrow^2-\zeta^2$.}.

\textit{Results: Collisionless Normal Gas} --- In
Fig.~\ref{spheric}, we present the monopole frequencies and
compare the results of the collisionless (full lines) with those
of the hydrodynamic approach for the normal shell. For both
approaches, the lowest monopole frequency is $2\omega_0$ for all
polarizations; this is in agreement with an exact result derived
by Castin~\cite{castin} and differs from the results of
Ref.~\cite{desilva2}. As expected for the collisionless case, the
second monopole frequency is $\sqrt{32/3}\omega_0$ at zero
polarization and $4\omega_0$ at full polarization. We find,
however, that the collisionless mode is damped for polarizations
between $0.09$ and $0.28$ (grey full line). The origin of this
damping is a resonance which occurs when $\tau=2\pi/\omega$; that
is, when there exist particles for which the time to travel from
the interface and back equals the period of the collective motion.
Mathematically, it corresponds to the pole in $\nu$ (see
Eqn.~\eqref{eq:soln}). The damping frequency is obtained by
writing $\omega\rightarrow\omega-i\Gamma$ with $\omega$ and
$\Gamma$ real and positive, and analytically continuing the
pressure tensor $\Pi_{rr}$ from negative to positive values of
$\Gamma$~\cite{landau,lazarides}. As is shown in the inset of
Fig.~\ref{spheric}, the maximal damping for this number of
particles is $\Gamma=0.05\omega_0$, which is weak but
experimentally detectable. We stress that the origin of this
damping is purely geometrical, and its presence and strength will
strongly depend on the trapping geometry.

\textit{Conclusion} --- We have established the boundary
conditions at the interface between the SF core and the
surrounding N shell for a trapped imbalanced fermion gas in an
isotropic trap at ultralow temperatures. Using these conditions we
have obtained the frequencies of collective modes, as a function
of the polarization, when the normal gas behaves hydrodynamically.
We have also calculated the frequencies of the monopole mode
for a collisionless normal gas.

For the hydrodynamic case, we find collective modes analogous to
those in a single-component system, but with shifted frequencies.
In addition, there exists a new class of collective modes; some of
these have energies lower than the trap frequency. These modes
correspond to out-of-phase motion of the SF core and the
surrounding N shell.

For the case of a collisionless normal gas, we find that the
lowest monopole mode remains at $\omega=2\omega_0$, in agreement
with an exact result due to Castin. We have also calculated the
next lowest mode, finding that it reduces to the appropriate
limits for vanishing and complete polarization of the trap and,
interestingly, that it is damped for some polarizations. This
damping is a geometric effect. It would be interesting to study
crossover behavior of the OOP modes from the hydrodynamic to the
collisionless normal gas.

\textit{Acknowledgements} --- We acknowledge partial support by Project No.~FWO
G.0115.06; B.V.S.~and A.L.~are supported by Project No.~GOA/2004/02. We thank
Joseph Indekeu for a careful reading of the manuscript and Theja De Silva for
bringing Ref.~\cite{castin} to our attention. We also thank Bidzina
Shergelashvili for useful discussions.

\end{document}